\documentclass [10pt]{article}
\usepackage[T1]{fontenc}
\usepackage[final]{epsfig}
\usepackage{amssymb}
\usepackage{multicol}
\usepackage{graphics}

\newcommand{\R}{\mathbb{R}}
\newcommand{\Ham}{{\mathcal H}}
\newcommand{\Z}{{\mathcal Z}}
\newcommand{\Zred}{{\mathcal Z_{N}}}
\newcommand{\IM}{{\mathtt i}}

\begin{document}

\title{Thermal-equilibrium properties of one-dimensional gas with short-ranged repulsion}

\author{\textbf{Milan Krbalek}\\
Faculty of Nuclear Sciences and Physical Engineering,\\
Czech Technical University,\\  Trojanova 13, 120 00 Prague, Czech
Republic\footnote{Electronic address:
\texttt{milan.krbalek@fjfi.cvut.cz}}}

\maketitle

\begin{abstract}
We derive the exact formula for thermal-equilibrium spacing
distribution of one-dimensional particle gas with repulsive
potential depending on the distance $r$ between the neighboring
particles. We are focused on the power-law potentials
$r^{-\alpha}$ for $\alpha>0,$ being motivated by the actual
traffic research.\\


\end{abstract}

%\section{Introduction}

\begin{multicols}{2}

Investigation of one-dimensional particle ensembles seems actually
to be very useful for understanding of the complex system called
\emph{vehicular traffic.} Beside the popular cellular automata, in
the recent time new trend appears in the traffic modelling.
Application of the equilibrium statistical-physics to
non-equilibrium traffic systems has been successfully demonstrated
for example in the Ref.\cite{Helbing_and_Krbalek}. In this
continuing work we analyze a certain family of the particle gases
exposed to the heat bath with the temperature $T \geq 0,$ being
primarily aimed at the equilibrium distribution for particle
distance (\emph{spacing distribution}) and particle velocity
(\emph{velocity distribution}). Subsequently, such a detailed
description can be used for comparing with the microscopic
structure of the realistic traffic flows (in
Ref.\cite{Helbing_and_Krbalek}).\\

%\section{Equilibrium distributions of the particle spacings}

Consider $N$ identical particles on the circle of the
circumference $L=N.$ Let $x_i$ $(i=1,2,\ldots,N)$ represents the
circular position of the $i-$th particle. Put $x_{N+1}=x_1+2\pi,$
for convenience. Now we introduce the short-ranged potential
energy
$$U \propto \sum_{i=1}^N V\left(r_i\right),$$
where $V(r_i)$ corresponds to the repulsive two-body potential
depending on the distance $r_i=|x_{i+1}-x_i|\frac{N}{2\pi}$
between the neighboring particles only. Thus, the hamiltonian of
the described ensemble reads as
$$\Ham=\frac{1}{2}\sum_{i=1}^N (v_i-\overline{v})^2+C\sum_{i=1}^N
V\left(r_i\right),$$
with the $i-$th particle velocity $v_i$ and the positive constant
$C.$ Note that $\overline{v}$ represents the mean velocity in the
ensemble. Then, the appropriate partition function\footnote{$k$
represents the Boltzmann factor}
\begin{equation}
\Z = \int_{\R^{2N}} \delta \left(L-\sum_{i=1}^N r_i\right)
e^{-\frac{\Ham}{kT}}~ dr_1 dv_1 \ldots dr_N  dv_N
\label{Partition_function01}
\end{equation}
leads us to the simple assertion that velocity $v$ of particles is
Gaussian distributed, i.e.
$$P(v)=\frac{1}{\sqrt{2\pi k T}}~ e^{\frac{(v-\overline{v})^2}{2kT}}.$$
is the corresponding probability density.\\

Of larger interest, however, is the spacing distribution
$P_\beta(r)$. In order to calculate the exact form of
$P_\beta(r)$, one can restrict the partition function
(\ref{Partition_function01}) to the reduced form
$$\Zred(L)=\int_{\R^N} \delta \left( L-\sum_{i=1}^N r_i\right) e^{-\beta\sum_{i=1}^N V\left(r_i\right)}\,dr_1\ldots\,dr_N$$
where $\beta=\frac{C}{kT}$ is so-called \emph{inverse temperature}
of the heat bath. Denoting $f(r)=e^{-\beta V(r)}$ the previous
expression changes to
$$\Zred(L)=\int_{\R^N} \delta \left( L-\sum_{i=1}^N r_i\right) \prod_{i=1}^N f(r_i)~dr_1\ldots\,dr_N.$$
Applying the Laplace transformation (see the Ref.\cite{Bogomolny}
for details) one can obtain
$$g_N(p) \equiv \int_0^\infty \Zred(L)~ e^{-pL}\,dL=$$

$$=\left( \int_0^\infty f(r) e^{-pr}\,dr \right)^N \equiv \left[
g(p) \right]^N.$$
Then the partition function (in the large $N$ limit) can be
computed with the help of Laplace inversion
$$\Zred(L)=\frac{1}{2\pi\IM} \int_{B-\IM\infty}^{B+\IM\infty}
g_N(p) ~ e^{Lp}~dp.$$
Its value is well estimated by the approximation in the saddle
point $B$ which is determined using the equation
$$\frac{1}{g(B)}\frac{\partial g}{\partial p}(B)=-\frac{L}{N}.$$
Thus,
\begin{equation} \Zred(L) \approx \left[ g(B) \right]^N e^{LB}.
\label{party_sum} \end{equation}
Hence the probability density for spacing $r_1$ between the
particles $\sharp 2$ and $\sharp 1$ can be then reduced to the
form
$$P(r_1)=\frac{\Z_{N-1}(L-r_1)}{\Zred(L)} f(r_1).$$
Supposing $N \gg 1$ and using equation (\ref{party_sum}) we obtain
$$P(r_1)=\frac{1}{g(c)} f(r_1) ~ e^{-Br_1},$$
which leads to the distribution function for spacing $r$ between
arbitrary couple of neighboring particles
\begin{equation}
P_\beta(r)=A ~ e^{-\beta V(r)} ~ e^{-Br}. \label{LS}
\end{equation}
Note that constant $A$ assures the normalization $\int_0^\infty
P_\beta(r)~dr=1.$ Furthermore, returning to the original choice
$L=N,$ the mean spacing is
\begin{equation}
\langle r \rangle \equiv \int_0^\infty r P_\beta(r)~dr=1.
\label{Scaling}
\end{equation}
%
%\section{Normalization of the spacing distribution}
%
Two above conditions can be understood as equation system for
unknown normalization constants $A,B.$\\

Let us to proceed to the special variants of the gas studied.
Firstly, we draw our attention to the Coulomb gas with the
logarithmic potential

$$V(r):=-\ln(r).$$
Such a gas is frequently used in the many branches of physics
(including the traffic research in Ref. \cite{Wagner}) and the
corresponding spacing distribution reads as (see Ref.
\cite{Bogomolny})

$$P_\beta(r)=\frac{(\beta+1)^{\beta+1}}{\Gamma(\beta+1)}~r^\beta
e^{-(\beta+1)r},$$
where $\Gamma(\xi)$ is gamma function. Of larger physical
interest, as demonstrated in Ref. \cite{Helbing_and_Krbalek} and
\cite{Seba}, seems actually to be the potential

$$V_\alpha(r):=r^{-\alpha},$$
for $\alpha>0.$ The aim of the following computational procedure
is to normalize the distribution

\begin{equation}
P_\beta(r)=Ae^{-\frac{\beta}{r^\alpha}}e^{-Br}. \label{spacing
distribution}
\end{equation}
Consider now the favorable choice $\alpha=1,$ for which the
normalization integrals are exactly expressed as

\begin{eqnarray}
\int_0^\infty e^{-\frac{\beta}{r}} e^{-Br} ~dr= 2
\sqrt{\frac{\beta}{B}}~ \mathcal{K}_1\left(2\sqrt{\beta B} \right)
\nonumber \\ \int_0^\infty
re^{-\frac{\beta}{r}}e^{-Br}~dr=2\frac{\beta}{B}~\mathcal{K}_2\left(2\sqrt{\beta
B} \right), \label{label}
\end{eqnarray}
where ${\mathcal{K}}_\lambda$ is the Mac-Donald's function
(modified Bessel's function of the second kind) of order
$\lambda,$ having for $\lambda=1$ and $\lambda=2$ an approximate
expression

$$\mathcal{K}_\lambda(y) = \sqrt{\frac{\pi}{2}}e^{-y}
\left(y^{-1/2} + \frac{3}{8} 5^{\lambda-1} y^{-3/2} +
\mathcal{O}\left(y^{-5/2}\right)\right).$$
Applying the equations (\ref{label}) to the normalization
integrals one can determine the exact values of the constants $A$
and $B$. Both of them can be, moreover, very well estimated by the
large $\beta$ approximations

$$B \approx \frac{900~
\beta}{\left(3-16\beta+\sqrt{9+864\beta+256\beta^2}\right)^2}
\approx \beta + \frac{3}{2}$$
and

$$A \approx \frac{1}{2} \sqrt{1+\frac{3}{2\beta}} \frac{1}{\mathcal{K}_1\left(\sqrt{2\beta
(2\beta+3)}\right)}.$$

Finally, we investigate the distribution (\ref{spacing
distribution}) for general $\alpha>0.$ Although in this case the
normalization integrals are not trivially solvable, the scaling
(\ref{Scaling}) leads us to the simple formula

\begin{equation} B \approx \alpha\beta+1 + \frac{\alpha}{2}\hspace{1cm} \left(\beta \rightarrow
\infty\right).\label{extim_B} \end{equation}
The rough estimation $r^{-\alpha} \approx 1 - \alpha + \alpha
r^{-1}$ provides the asymptotical formula for normalization
constant $A:$

\begin{equation} A \approx \frac{1}{2}
\sqrt{1+\frac{1}{2\beta}+\frac{1}{\alpha \beta}}
\frac{e^{\beta(1-\alpha)}}{\mathcal{K}_1\left(\sqrt{2\alpha\beta
(2\alpha\beta+\alpha+2)}\right)}. \label{extim_A} \end{equation}

\begin{center}
\scalebox{.4}{\includegraphics{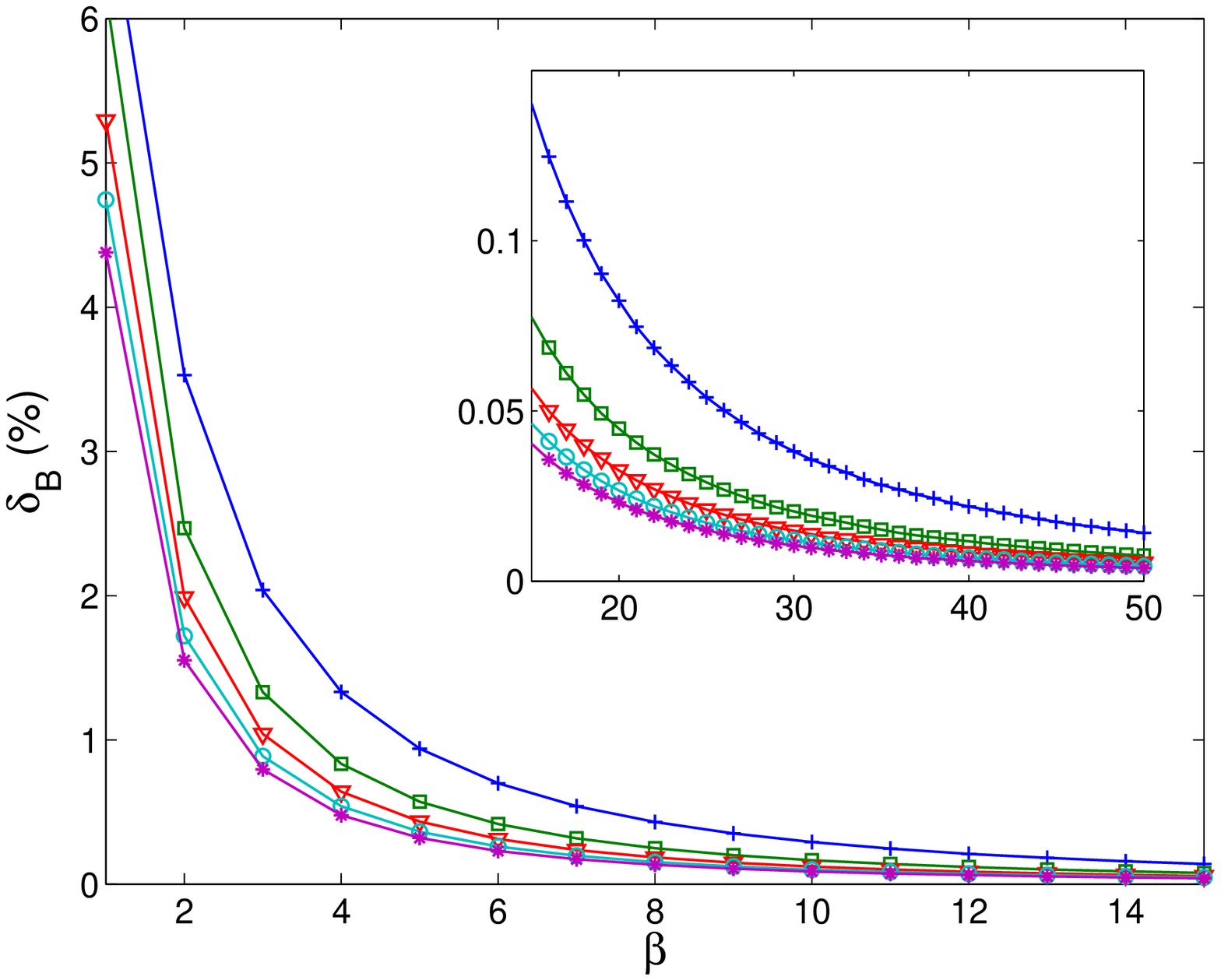}} \footnotesize

\textbf{Figure 1.} Relative deviation in the approximate value of
the
normalization constant $B$ as function of parameter $\beta.$\\
We display the deviation (\ref{disc_B}) between the numerical
value $B_{ex}$ and the value $B_{est},$ obtained from the large
$\beta$ approximation (\ref{extim_B}). The plus signs, squares,
triangles, circles and stars correspond to the parameters
$\alpha=1,2,3,4,5,$ respectively. The tails of the curves are
magnified on the insert.

\end{center}

For practical applications it seems to be useful to detect the
critical inverse temperature $\beta_{crit}$ under which the
relative deviation between the exact ($ex$) and estimated ($est$)
values of the constant A (or B)

\begin{equation}
\delta_{\log(A)}:=\frac{|\log(A_{ex)}-\log(A_{est})|}{\log(A_{ex})}
\label{disc_A} \end{equation}

\begin{equation} \delta_B:=\frac{|B_{ex}-B_{est}|}{B_{ex}} \label{disc_B} \end{equation}
are larger then the fixed permissible deviation $\delta.$  For
these purposes we plot the functional dependence
$\delta_B=\delta_B(\beta)$  and
$\delta_{\log(A)}=\delta_{\log(A)}(\beta)$ in the Figure 1 and
Figure 2, respectively. We note that the exact values
$A_{ex},B_{ex}$ were determined with the help of the numerical
computations.

\begin{center}
\scalebox{.4}{\includegraphics{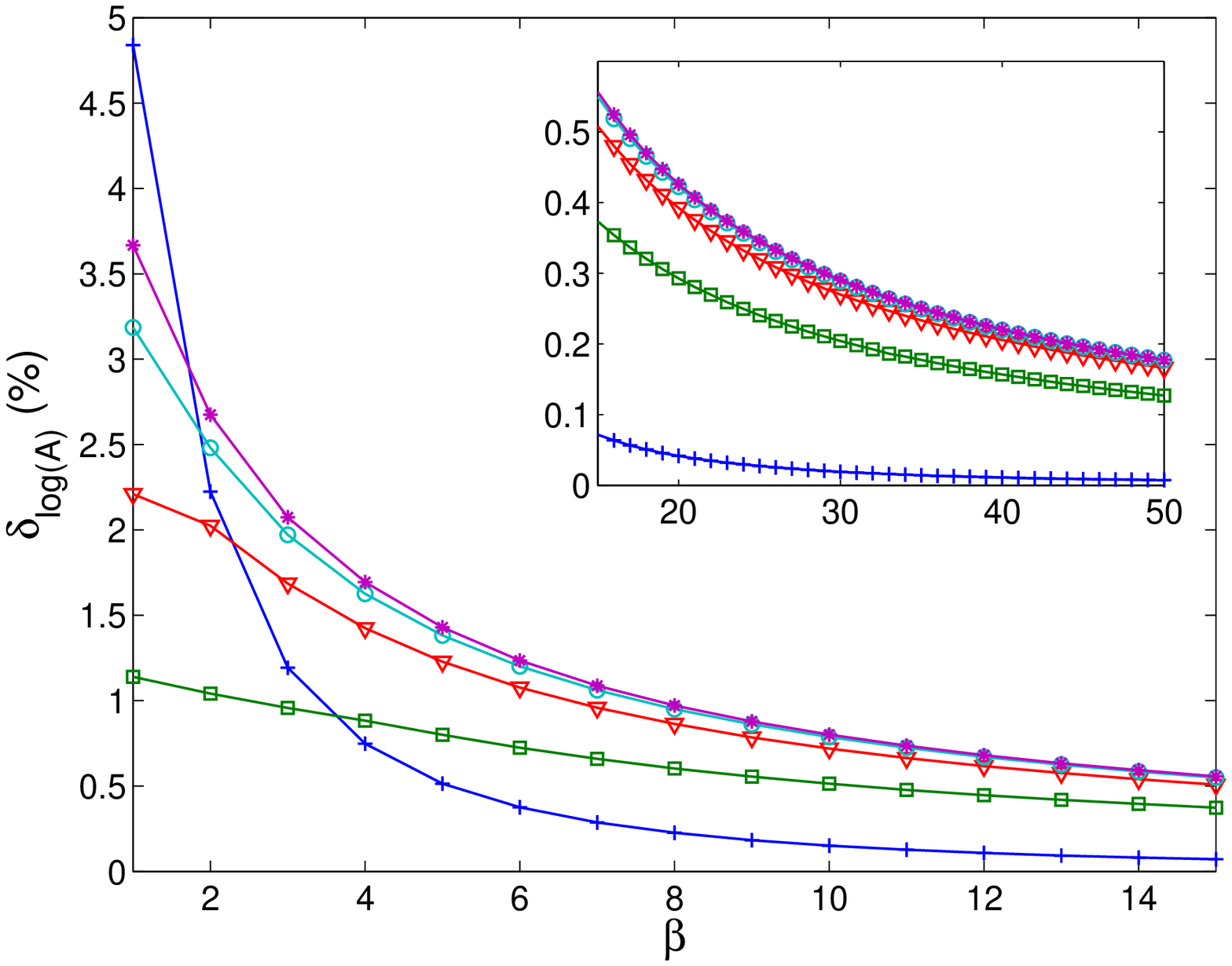}} \footnotesize

\textbf{Figure 2.} Relative deviation in the approximate value of
the
normalization constant $A$ as function of parameter $\beta.$\\
Plotted is the deviation (\ref{disc_A}) between the numerically
computed value $A_{ex}$ and the estimated value (\ref{extim_A}).
The symbols used here are consistent with the symbols in the Fig.
1.

\end{center}

To conclude, we have found the exact form of the
thermal-equilibrium spacing distribution for one-dimensional gas
which neighboring particles are repulsed by the two-body potential
$V=r^{-\alpha},$ where $r$ is their mutual distance. The values of
two normalization constants were successfully estimated by the
simple approximation. The determined distribution has been and
will be applied in the traffic research where compared to the
distance clearance distribution of the freeway traffic samples.

\end{multicols}

\end{document}